\documentclass[a4paper,fleqn,usenatbib]{mn2e}

\usepackage{url,ulem,times,graphicx,amsmath,amsfonts,amssymb,color,epsfig,epstopdf}
\usepackage{graphics}
\usepackage{epsf}
\usepackage{bm}
\usepackage{color}
\usepackage{rotating}
\usepackage[T1]{fontenc}
\usepackage{ae,aecompl}
\usepackage{array,multirow}
\usepackage[percent]{overpic}
\usepackage{amsfonts,amssymb,amsmath}
\usepackage{aas_macros}
\usepackage{varioref,textcomp}

\usepackage{hyperref}
\hypersetup{
   colorlinks=true,
   urlcolor=blue,
   citecolor=blue,
   linkcolor=blue,
   pdfborder= 0 0 0
}

\newcommand{\Rmnum}[1]{\uppercase\expandafter{\romannumeral #1}}

\title[Analytical model for the Milky Way satellites]
      {Warm dark matter model with a few keV mass is bad for the too-big-to-fail problem}
\author[X. Kang]
{Xi Kang$^{1,2}$ \thanks{E-mail:kangxi@pmo.ac.cn}\\
$^1$Purple Mountain Observatory, No 8 Yuanhua Road, Nanjing 210034, China \\
$^2$Zhejiang University-Purple Mountain Observatory Joint Research Center for Astronomy, Zhejiang University, Hangzhou 310027, China\\
}

\date{Accepted xxxx. Received xxxx; in original form xxxx}

\pubyear{2019}

\begin{document}
\label{firstpage}
\pagerange{\pageref{firstpage}--\pageref{lastpage}}
\maketitle

\begin{abstract}

Theoretical studying of the very inner structure of faint satellite galaxy requires very high-resolution hydro-dynamical simulations with realistic model for star formation, which are beginning to emerge recently. In this work we present an analytical description to model the inner kinematic of satellites in the Milky Way (MW). We use a Monte-Carlo method to produce merger trees for MW mass halo and analytical models to produce stellar mass in the satellite galaxies. We consider two important processes which can significantly modify the inner mass distribution in satellite galaxy. The first is baryonic feedback which can induce a flat inner profile depending on the star formation efficiency in the galaxy. The second is the tidal stripping to reduce and re-distribute the mass inside satellite. We apply this model to MW satellite galaxies in both CDM and thermal relic WDM models. It is found that tidal heating must be effective to produce a relatively flat distribution of the satellite circular velocities, to agree with the data. The constraint on WDM mass depends on the host halo mass. For a MW halo with dark matter mass lower than $2\times 10^{12}M_{\odot}$, a 2 keV WDM model can be safely excluded as the predicted satellite circular velocities are systematically lower than the data. For WDM with mass of 3.5 keV, it requires the MW halo mass to be larger than $1.5\times 10^{12}M_{\odot}$, otherwise the 3.5 Kev model can also be excluded. Our current model can not constrain the WDM model with mass larger than 10 Kev.


\end{abstract}




\section{Introduction}
\label{sec:intro}

The Milky Way (MW) is an ideal  local laboratory to test the nature of
dark matter and galaxy formation  physics, as its proximity allows for
accurate measurements  of the mass  and velocity distributions  of the
satellite galaxies.  The two well-known  challenges on the  cold dark
matter (CDM)  model are the  missing satellite problem which states that the CDM model predicts hundred  of subhaloes  but only  a dozen  classical satellite
galaxies are observed \citep{Klypin99}, and the too-big-to-fail problem
(TBTF) stating that the most massive subhaloes are  too centrally dense to host the
brightest  satellite galaxies  \citep{Boylan-kolchin11}. These  problems
have produced a larger number of studies in the past years. For a recent review
on this topic, we refer the readers to the paper by \cite{Bullock17}.

To rescue  the classical CDM  model, the  most common solution  is invoking baryonic effects.  It  was early proposed that  the cosmic
re-ionization  can  effectively  suppress   the  gas  accumulation  in
low-mass  halo \citep[e.g.,][]{Gnedin00}  and galaxy  formation models
which include  this effect can  well match the luminosity  and spatial
distribution    of    the    satellite     galaxies    in    the    MW
\citep[e.g.,][]{Maccio10a}.  On the other  hand, the baryonic feedback
from star formation can also  produce a core-like profile in satellite
galaxies     to     alleviate      the     TBTF    problem
\citep[e.g.,][]{Governato12,Brooks14}.  Other  dynamical effects, such
as MW disc heating and tidal  stripping, are also helpful to lower the
central density profile  of satellite galaxies and  some studies claim
such  an  effect is  more  efficient  in  satellites with  an  initial
shallower                                                      profile
\citep[e.g.,][]{Zolotov12,Ogiya15,Dutton16,Sawala16,Tomozeiu16}.

Along  with the  efforts  of  invoking baryonic  feedback  in CDM,  an
alternative solution is to change  the properties of dark matter, 
  such     as     the     self-interacting    cold     dark     matter
  \citep[e.g.,][]{Vogelsberger12},     the     warm    dark     matter
  \citep[e.g.,][]{Polisensky11}   or    the   combination    of   both
  \citep{Cyr-Racine16, Vogelsberger16}.  These  alternative dark matter
  models seem to be more promising  as they can reduce both the number
  of  subhaloes  and  their  central  density 
  \citep{Bode01, Colin00, Vogelsberger19}.
For the WDM model, constraints based on satellite counts  and kinematics are getting stronger. For example, using the  satellites count of the MW as  a constraint, a few
studies \citep{Maccio10b,  Polisensky11, Lovell14}  have found  that a
thermal relic warm  dark matter  candidate  with mass $m_{\nu}  >
1-2.3$ keV is  needed, as a lower mass candidate  will produce too few
satellite galaxies than observed.  Similar constraint is also obtained
by \cite{Kang13} using  a joint constraints from  local galaxy stellar
mass  function and  the  Tully-Fisher relation.   On  the other  hand,
\cite{Shao13} found that a WDM mass around $\sim 0.5$ keV is needed to
produce    a   density    core   consistent    with   the    satellite
kinematics. \cite{Lovell12} used N-body simulation  to show that a WDM
mass  with   $m_{\nu}  \sim  2$   keV  is  needed  to   alleviate  the
TBTF  problem.  Thus,  the  most possible  WDM mass  around
$\sim 2$ keV is required to simultaneously match the number counts and
kinematics  of  the  Milky  Way   satellites.   However,  this  is  in
disagreement with the constraint on the  WDM mass with $m_{\nu} > 3.3$
keV using Ly-$\alpha$ forest observations \citep[e.g.,][]{Viel13} though this  constraint is  affected by the  thermal history  of the
  intergalactic medium \citep{Garzilli17}.

Regardless of  the nature  of dark matter,  most studies  have reached
consensus that reproducing  the satellite luminosity distribution
  is key to the TBTF problem.  In particular, the degree to which haloes have shallower density profiles than expected in CDM or even a central core, if cores indeed exist in nature,  is  found  to be  strongly
correlated with the  star formation efficiency in  the satellite using
hydyodynamical     simulations    \citep[e.g.,][]{DiCintio14,Tollet16,
  Fitts17} while other  simulations suggest that core  creation is not
an  inevitable outcome  but depending  on star  formation history  and
subgrid physics  \citep{Grand17,Bose19}.  In addition, to  resolve the
very inner  kinematic structure  of satellites,  one need  to simulate
their evolution with very high  spatial- and mass resolution.  Putting
together,  it  requires  simulation to  simultaneously  reproduce  the
stellar mass and  kinematics of the MW satellites.  Such  an effort is
only  recently  achieved   from  the  state-of-the-art  hydrodynamical
simulations,    such   as    NIHAO   \citep{Wang15,Buck19},    APOSTLE
\citep{Sawala16}             and            FIRE-2             project
\citep{Wetzel16,Garrison-Kimmel19}.   These simulations  claimed
  that  the  two  problems  can  be  simultaneously  solved  and  they
  suggested that  tidal stripping and heating,  not baryonic feedback,
  is  the  main  process  leading  to  lower  velocity  dispersion  of
  satellites.  As  the   tidal  effects  depend  on   the  inner  mass
  distribution in satellite galaxy \citep[e.g.,][]{Penarrubia10} which
  is  therefore affected  by baryonic  feedback, one  needs controlled
  simulations to separate  the baryonic feedback and  tidal process so
  as to claim which one is the dominant.

Undoubtedly, hydrodynamical simulation is the best tool to address the
satellite problems.  However, the high computational cost has hindered
the accomplishment  of a large sample  of MW analogous to  capture the
formation history of  the MW. Given the limited  realistic samples and
the  uncertainty  in subgrid  physics  in  the present  hydrodynamical
simulations, it is still not clear  what is the main physical process,
either feedback  induced core or  tidal stripping, to account  for the
lower  mass density  in  observed satellite  galaxies.   To this  end,
analytical model,  which can  easily follow  the formation  history of
large sample of MW analogous and include relevant baryonic process, is
particularly desired.  In this work, we  aim to construct such a model
based on current results.  We then apply the model to both CDM and WDM
models to investigate the separate effect of feedback induced core and
tidal stripping on  the satellite kinematics, and  in particularly, to
set constraints on the mass of the thermal relic WDM model.

Our  analytical model  contains the  following basic  ingredients.  To
model  the  formation  history  of  the  MW  type  halo,  we  use  the
Monte-Carlo code  by \cite{Parkinson08} to  produce a large  sample of
merger trees  for MW  analogs. To  set the  stellar mass  of satellite
galaxies, we use either the  semi-analytical model \citep{Kang12} or a
simple  abundance matching  method.  To  model the  kinematics of  the
satellite galaxies,  we follow three  steps.  Firstly, we  assume that
before satellite is accreted, its  DM distribution initially follows a
NFW profile.   Secondly, we  modify the inner  slope of  the satellite
using  the fitted  relation between  the  density slope  and the  star
formation efficiency in  satellite galaxy \citep[e.g.,][]{DiCintio14}.
Thirdly,  after  satellite  is  accreted by  the  host  galaxy,  tidal
stripping from host galaxy will strip the DM mass of the satellite and
the  associated  tidal   heating  will  also  alter   its  inner  mass
distribution. To model this process, we  use the model for the average
mass loss of  subhalo by \cite{Giocoli08} and use  the fitting formula
by \cite{Penarrubia10} to  model the effect of tidal  stripping on the
inner density structure of the satellite. With the above procedure, we
are then  able to predict  the circular  velocity of satellite  at any
given radius and compare the model predictions to the data.

Our model is  in spirit similar to the recent  work by \cite{Lovell17}
in  which they  also use  analytical  model to  address the  satellite
kinematics, particularly  in the  WDM model.   Our model  is different
from  theirs in  many details.  For example,  they use  the simulation
results  of \cite{Sawala16}  to  include the  baryonic  effect on  the
maximum  circular velocity $V_{max}$ of  the  satellite galaxy,  and
compare the  distribution of $V_{max}$  with that obtained for  the MW
satellites.  In our  model, the  baryonic effect  is reflected  in the
change of the  inner density profile which depends on satellite
star  formation  efficiency  \citep[e.g.,][]{DiCintio14,Fitts17}.   In
fact,  the  baryonic  effect  in  these  based  simulations  are  very
different  and  there  is  no core  creation  in  the  \cite{Sawala16}
simulation.  Furthermore,  $V_{max}$ is not directly  observed for the
satellites and there  is significant scatter on  the derived $V_{max}$
from the  observed circular velocities \citep{Sawala16}.  While in our
model we  directly compare the  circular velocity  to the data  and it
requires us to model the inner  slope of the satellites.  In addition,
the modelling of  tidal effect on satellite mass  distribution is also
different between the two models.

The  paper  is  organized  as follows.   In  Sec.~\ref{sec:method}  we
introduce  our  analytical method  including  the  production of  halo
merger  trees, method  to  assign density  profile  of subhalo  before
infall, analytical model of star  formation for the satellite galaxies
and the modification  of their inner slopes by  supernova feedback and
tidal  effect.    We  present   the  comparison   with  the   data  in
Sec.~\ref{sec:results}   and  conclude   with   short  discussion   in
Sec.~\ref{sec:discuss}.


\section{Analytical Methods}
\label{sec:method}

Modelling  the properties  of observed  satellite galaxies  in the  MW
requires accurate information of the formation history of the MW, such
as the mass  growth of the host halo,  the infall time and  orbits of the satellite galaxies. Current studies have been trying to derive these information \citep[e.g.,][]{Rocha12,Simon18}, but accurate recovery of the infall history of the satellites  is still challenging. So we
produce a large sample of MW-mass halo around $10^{12}M_{\odot}$ using
Monte-Carlo based method. We then model the formation and evolution of
luminous satellite galaxy using an  analytical model or using a simple
abundance matching method  to assign stellar mass to  the subhalo. For
each observed  satellite galaxy, we  select its
counterparts from  our model with similar  luminosity or
stellar  mass, and  compare the  observed properties  of the  satellite
galaxy to the statistical distribution of these model satellites.

\subsection{merger tree of MW type halo}

We use the Monte-Carlo merger tree code from \cite{Parkinson08} to produce the formation history of MW-type halo with mass of $10^{12}M_{\odot}$ in both the CDM and WDM models. This algorithm is based on the extended Press-Schechter formalism and revised to match the halo merger tree from Millennium Simulation \citep{Springel05}. \cite{Cole08} have shown the improvement from this Monte-Carlo algorithm. The basic input of this method is the analytical power spectrum \citep[e.g.,][]{Bardeen86} and the cosmological parameter for which we use the WMAP9 cosmology \citep{Hinshaw13}. We produce 5000 realizations of the merger tree for the MW-type halo. 

In the  WDM case, the velocity  dispersion of DM  particles produces a
characteristic  free-streaming  scale below  which  the structure  is
suppressed.  The   lighter  the  particle  mass  is,   the  larger  the
free-streaming  scale. The  impact  of free  streaming  on the  power
spectrum is  to modify  the transfer function  in WDM as  suggested in
\cite{Bode01},
\begin{equation}
T(k) = (\frac {P_{Lin}^{WDM}} {P_{Lin}^{CDM}})^{1/2} = (1+(\alpha k)^{2\mu})^{-5/\mu}
\label{eq:tk}
\end{equation}
where  $P_{Lin}^{WDM}$  and   $P_{Lin}^{CDM}$  are  the  linear  power
spectrums in  WDM and CDM  model. \cite{Viel05}  obtained $\mu=1.2$
and found that $\alpha$ is related  to the WDM mass, $m_{\nu}$, as the
following,
\begin{equation}
\alpha = 0.049(\frac {m_{\nu}} {1keV})^{-1.11}(\frac {\Omega_{\nu}} {0.25})^{0.11}(\frac {h}{0.7})^{1.22}h^{-1}Mpc,
\label{eq:alphawdm}
\end{equation}
We use  the above equations to  derive the initial power  spectrum for
the WDM model.  As  noted by \cite{Kennedy14},  also see \cite{Benson13}, the usually adopted top-hat filter is not applicable in the
presence of a  cutoff in the power  spectrum, as was in  the WDM case,
and a sharp  filter in $k$-space is used instead.  We refer the readers to the \cite{Kennedy14} paper  for detail. In this work, we  consider a few WDM
mass,  with  $m_{\nu}=1,2,3.5,10$ keV.  For  each  WDM mass,  we  also
generate  5000 merger  trees  for  the MW-type  host  halo  with $M_{vir}  =
10^{12}M_{\odot}$.

\begin{figure}
\centerline{\epsfig{figure=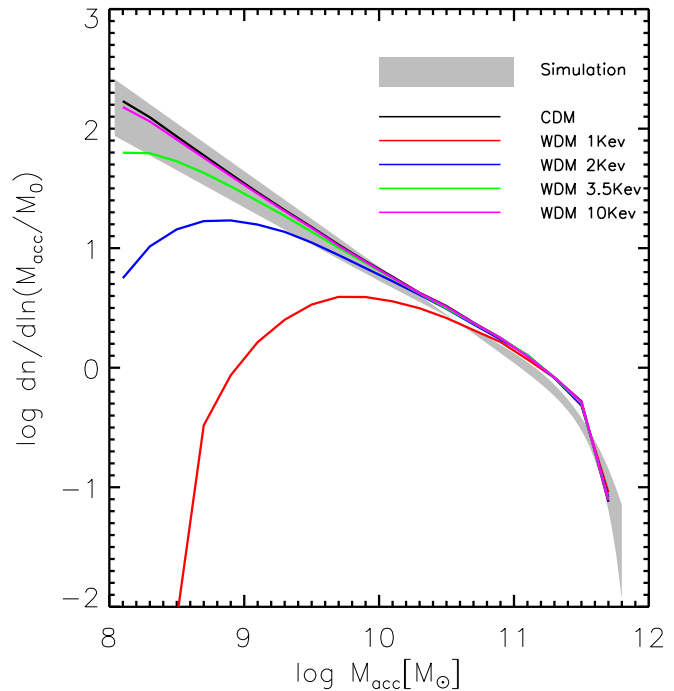,width=0.49\textwidth}}
\caption{The mass function of accreted subhaloes by a Milky Type host halo with mass around $10^{12}M_{\odot}$. The black line is for CDM and color lines are for the WDM models using the transfer function (Eq.~\ref{eq:tk} and Eq.~\ref{eq:alphawdm}) with different mass. The hatched gray regions show the results from CDM simulations.}
\label{fig:massfn_acc}
\end{figure}

To see the effect of suppressed  power spectrum at small scales on the
satellite galaxy population in the WDM  model, it is firstly useful to
see on  which scale the  formation of  low-mass haloes is  reduced. In
Fig.~\ref{fig:massfn_acc},  we show  the  mass  functions of  accreted
subhaloes by the host halo during its formation history in the CDM and
WDM  models   using  the  power  spectrum   from  Eq.~\ref{eq:tk}  and
Eq.~\ref{eq:alphawdm}.  The  hatched  gray  region   shows  the
  simulation results  from \cite{Gao04}  and \cite{Stewart08}  for CDM
  model. It is seen that in  the WDM model, the formation of low-mass
haloes is suppressed and the extent of suppression is dependent on the
particle mass.  One  direct consequence of this result is  that, for a
fixed star formation efficiency (stellar  mass to halo mass relation),
the  number  of   luminous  satellites  would  be  less   in  the  WDM
model. However, given the current uncertainty in star formation model,
we will  try to  match the observed  satellite luminosity  function by
tuning the free parameters in the galaxy formation model as introduced
below.  Even so,  we will find that the WDM  model with lower particle
mass  ($m_{\nu}<  3.5$  keV)  is difficult  to  produce  enough  faint
satellite galaxies.

\subsection{Model for galaxy formation}
\label{sec:modelstar}

One key ingredient of our model is to set the inner structure of the satellites. As previous introduced, the inner density profile of satellite galaxy will be affected by the baryonic feedback.  To this end, we firstly need to determine both dark matter halo mass and the stellar mass of the satellite galaxies. We use two methods to assign stellar mass to dark matter subhalo. The first one is the semi-analytical  model (SAM) of \cite{Kang12} which is an updated version of our previous model \citep{Kang05,Kang09}. The SAM  includes key processes governing galaxy formation, such as gas cooling,  star formation, supernova feedback. We refer the readers to the paper by \cite{Kang05} for details. In \cite{Kang09} we included a model for the effect of cosmic reionization on the hot gas content of low-mass haloes \citep[e.g.,][]{Gnedin00}, by using a  filtering mass which corresponds to a mass scale at which haloes will only be able to
accrete  half of  the  universal baryonic  fraction. The fraction  of
baryons that can be accreted as hot gas is parameterized as,
\begin{equation}
f_{b,acc}(z,M_{vir}) = \frac {f_b} {[1+0.26M_{F}(z)/M_{vir}]^{3}}
\end{equation}
where  $f_b$ is  the universal  baryon fraction  and $M_{vir}$  is the
virial mass  of the halo, and  the filter mass $M_{F}(z)$  is given by \cite{Kravtsov04}. For both the  CDM and WDM models,  we tune the
model  parameters, such  as  star formation  and  supernova
feedback efficiency, to best  match the stellar luminosity function of
satellites in  the MW \citep{Koposov08}. The results will be shown in Sec.~\ref{sec:results}. By doing so,  we can get
the stellar  mass, $M_{\ast}$, for each satellite as well as  its virial mass $M_{vir}$ at accretion time. 

Another method to select satellite galaxy is similar to the halo abundance matching method \citep[e.g.,][]{Vale04}. However, many studies \citep[e.g.,][]{Guo15,Brook15,Errani18} have shown that for satellites of the MW, there is a large scatter between subhalo mass and stellar mass, mainly due to the effect of cosmic reionization on the gas content of low-mass haloes, thus the most luminous satellites do not often live in the most massive subhaloes. However, the TBTF problem concerns more about the kinematic match between the simulated massive subhaloes with the observed luminous satellites. Thus following the spirit of abundance matching (AM) method, we take the nine subhaloes with most massive mass at accretion, but assign the observed stellar mass of the MW satellites \citep[e.g.,][]{Misgeld11} to these subhaloes, by putting the most luminous satellites to the most massive subhaloes at accretion. We have also tested that our results are not significantly affected if we instead use the highest maximum velocity of subhalo at infall. Given the stellar mass, halo mass and infall times of these satellites, we can use the following descriptions to determine their inner structures and circular velocities.

\subsection{Halo density profile and baryonic effect}

\begin{figure*}
\centerline{\epsfig{figure=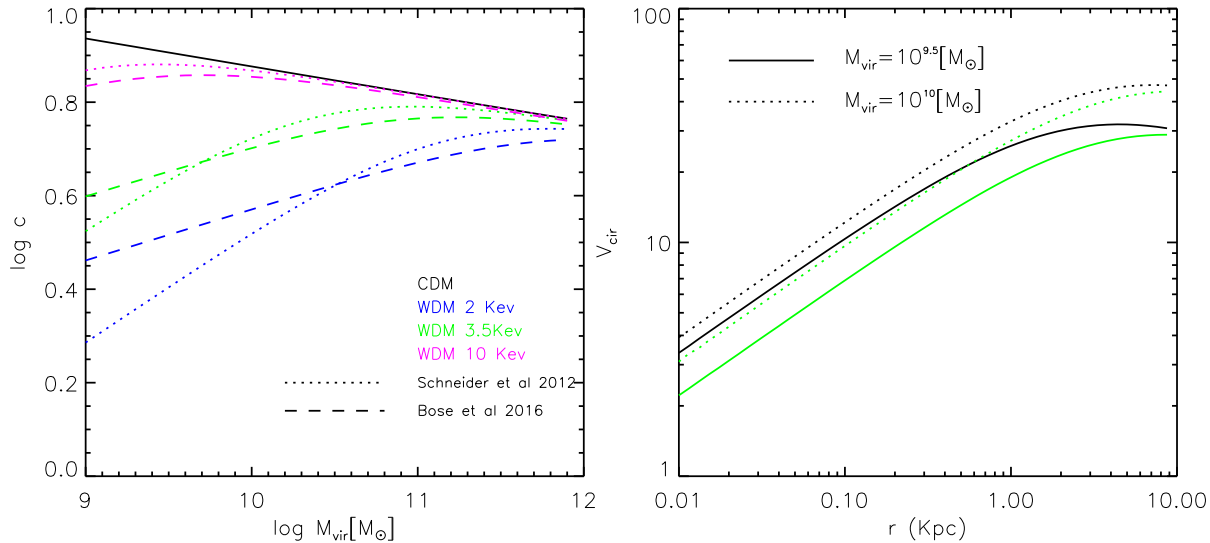,width=0.9\textwidth}}
\caption{Left panel: the halo concentration-mass relations at z=1 for the CDM and WDM models. The black line is the fitting formula from Prada et al. (2012) for CDM. The color lines are for WDM models with $c-M$ relation from Schneider et al. (2012) and Bose et al. (2016), respectively. Right panel: the circular velocity of two NFW haloes at z=1 in the CDM (black lines) and WDM with mass $m_{\nu} = 3.5$ keV.}
\label{fig:cmrelation}
\end{figure*}

To predict the circular velocity  of satellite galaxy at given radius, we need to specify its density profile. We consider two main process which will modify the mass and dynamical structure of the satellite galaxy: baryonic feedback and tidal stripping. We assume that before satellite is accreted, the main process in play is bryonic feedback which acts to change the inner profile of the galaxy. After accretion, the tidal force from the host halo will strip the DM mass of the satellite and the associated tidal heating will also continuously re-distribute the mass inside the satellite. In this section we introduce how to include the first effect. We assume that before accretion,  the dark matter halo of satellite initially follows a general  NFW profile  or  the so-called
$\alpha-$model \citep{Ogiya15},
\begin{equation}
\rho(r) = \frac {\rho_{0}r_{0}^{3}} {r^{\alpha}(r+r_{0})^{3-\alpha}}
\label{eq:rhor}
\end{equation}
where $\alpha,  \rho_{0}$ and $r_{0}$ are the  logarithmic slope of the
central density, the  scale  density and  scale  length.   In  this
description, $\alpha =  1$ corresponds to the classical NFW  profile \citep{Navarro97}, and $\alpha = 0  $ corresponds to a central core. Numerous studies have been devoted to fit $\alpha$ in dark matter only simulations, which is found to be between $1$ and $1.5$  \citep[e.g.,][]{Moore99,Jing00,Diemand08}. Using more high-resolution simulations \citep[e.g.,][]{Springel08}, it is shown that the halo mass density is better described by the \cite{Einasto65} profile with the power index being a function of radius, but NFW still gives a reasonable description to the simulation result. In our  calculation we assume that, before invoking baryonic feedback, all haloes have  an initial slope with $\alpha = 1$.

To predict  the circular velocity  using Eq.~\ref{eq:rhor}, one  needs to
specify the  halo concentration, $c  = R_{vir}/r_{0}$. In the  CDM model, halo  concentration,  $c_{CDM}$,  is  dependent  on  halo  mass  and
formation time  \citep[e.g.,][]{Bullock01, Zhao03} and a  few fitting  formulae have been provided to describe the  $c-M$ relations from simulations \citep{Bullock01, Maccio08,Dutton14,Diemer15}. Here we use  the $c-M$ relation
from \cite{Prada12}  to set the concentration for the satellite
galaxies using their virial  mass $M_{vir}(z_{acc})$ at accretion time
$z_{acc}$.  For the  WDM model, it is found that  the halo profile can
still  be  described  by the  NFW  profile  \citep[e.g.,][]{Eke01,Schneider12,Maccio13,Lovell14,Schneider15,Ludlow16}, but the concentration, $c_{WDM}$,
is   generally  reduced,   and   the   $c_{WDM}-M$   relation   is also 
non-monotonic, reaching  a peak at a mass  scale indicated by the
truncation scale  and decreasing  at higher
and lower  masses. Here we use  the fitting formula  from \cite{Schneider12} to describe the connection between the concentration $c_{CDM}$
in CDM and $c_{WDM}$ in WDM models,
\begin{equation}
\frac {c_{WDM}(M)} {c_{CDM}(M)} = (1+15\frac{M_{hm}} {M})^{-0.3}
\label{eq:cWDM}
\end{equation}
where  $M_{hm}$ is the  half-mode mass  which is  related to  the half
radius,       $\lambda_{hm}$,      as       $M_{hm}       =      \frac
{4\pi}{3}\bar{\rho}(\lambda_{hm}/2)^{3}$,  and  $\lambda_{hm}$ is  the 
length  scale at  which the  amplitude  of the  WDM transfer  function
(Eq.~\ref{eq:tk}) is  reduced to $1/2$.  A similar fitting formula for the $c-M$ relation in WDM is also provided in \cite{Bose16}.

As an illustration, in left panel of Fig.~\ref{fig:cmrelation} we show the $c-M$ relation at z=1 for the CDM and WDM models. The colorful lines are predictions for WDM models. It is found that the fitting formulae from \cite{Schneider12} and \cite{Bose16} are very similar and their difference is less than 20\% at $M_{vir} > 10^{9}M_{\odot}$. For the following analysis we use the $c-M$ relation from \cite{Schneider12}. Compared with the $c-M$ relation in the CDM model, this relation at the low mass end in WDM is steeper and halo concentration is significantly decreased, depending on the WDM mass. At halo mass around $10^{10}M_{\odot}$, the concentration in WDM with $m_{\nu}=3.5$ keV is around 63\% of that in the CDM model. This decrease in $c$ will result in a lower circular velocity, as shown in the right panel where we plot the circular velocity of two haloes at z=1 in both CDM and WDM model with $m_{\nu}=3.5$ keV, respectively. This plot shows that in the CDM model halo circular velocity depends weakly on halo mass as the $c-M$ relation is relatively flat, but in the WDM model the halo mass dependence is stronger mainly due to the steeper $c-M$ relation at the low-mass end.

\begin{figure}
\centerline{\epsfig{figure=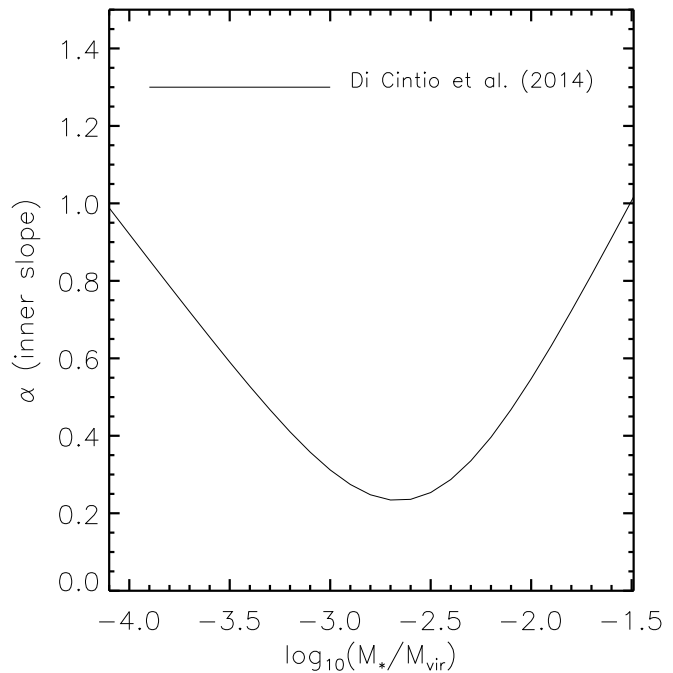,width=0.49\textwidth}}
\caption{The impact of baryonic feedback on the inner dark matter density profile,$\alpha$, as a function as the star formation efficiency, $M_{\ast}/M_{vir}$. The solid line is the fitting formula from Di Cintio et al. (2014)} 
\label{fig:fittingformula}
\end{figure}

To model the baryonic feedback on the inner density slope of dark matter in satellite galaxy,  we  use the  fitting  formula by  \cite{DiCintio14} which is only dependent on the star formation efficiency,
\begin{equation}
\alpha(x)  = -0.06  -  \log_{10}[(10^{x+2.56})^{-0.68}+(10^{x+2.56})]
\label{eq:alpha}
\end{equation}
 where $x = log_{10}(M_{\ast}/M_{vir})$, is the star formation efficiency given from  the semi-analytical  model or the halo AM method. Similar results have been obtained from other state-of-art  hydro-dynamical simulations  \citep[e.g.,][]{Tollet16,Fitts17,Hopkins18}. To simply illustrate  the dependence of
 $\alpha$ on the star formation efficiency in the galaxy, we plot this
 fitting formula in Fig.~\ref{fig:fittingformula}. It  is seen  that the  baryonic effect  peaks at
 around  $M_{\ast}/M_{vir} \sim  0.005$ where  the feedback  energy is
 enough to redistribute the inner dark matter and create a flat profile.  For low  star  formation efficiency  with $M_{\ast}/M_{vir}  <
 10^{-4}$, the feedback energy is  not enough to expel the dark matter
 distribution. For high star formation with $M_{\ast}/M_{vir} > 0.03$,
 the gravity from the excess stars in galaxy center will drag dark matter in and create a steeper profile.

\subsection{Tidal effect on satellite mass and inner profile} 

The second  process included  in  our  model  is  the effect of tidal stripping  on
satellite inner  structure. It  has long  been recognized  that, after
accretion, the strong tidal force from the host halo will strip the DM
mass  of the  satellites and the associated tidal heating (or tidal stirring)  will change  their inner
structure,  leading to  a decrease  of dark  matter mass  and velocity
dispersion \citep[e.g.,][]{Gnedin99}.  Earlier  studies \citep[e.g.,][]{Read06} claimed that the tidal effect  is not strong enough to bring the
satellites with  initial cusp profile  to match the data.  It is
later found that  the tidal effect is more evident  in satellites with
initial  flat slope  and it  can even  lead to  a total  disruption of
satellite  galaxy  \citep[e.g.,][]{Penarrubia10}.   More  recent
simulations  have  shown that  a  combined  effect from  initial flat 
density profile and tidal heating can lower the circular velocity of
satellites to agree with the data \citep[e.g.,][]{Brooks14,Tomozeiu16,Frings17}. 

The tidal  effect depends  on  the orbit  of satellite and it is difficult to estimate the accurate degree of tidal reduction on satellites  mass without exact information of its accretion  time and infall orbit. Fortunately,  in the  simulation by \cite{Penarrubia10}  they found that for satellite  with given
slope ($\alpha$), the  evolution of the structure  parameters, such as
$r_{max}$, $V_{max}$, can be better described solely by the amount of stripped
dark  matter  mass.  For  halo   with  initial  profile  described  by
Eq.~\ref{eq:rhor},  \cite{Penarrubia10}  provided  a  simple
empirical formula to fit the evolution of the structure parameters as,
\begin{equation}
g(x) = \frac {2^{\mu}x^{\eta}}{(1+x)^{\mu}}
\label{eq:gx}
\end{equation}
where $g(x)$ presents either $V_{max}/V_{max,0}$ or $r_{max}/r_{max,0}$, and $x=m_{s}/m_{s,0}$. Here $m_{s,0}, V_{max,0}, r_{max,0}$ are the halo mass, maximum circular velocity, corresponding radius of $V_{max,0}$ of satellites before accretion. \cite{Penarrubia10} listed a few fitted values of $\mu(\alpha)$ and $\eta(\alpha)$ for haloes with initial profile $\alpha=0.0$, $0.5$ and $1.0$. In the recent work by \cite{Lovell17} and \cite{Hiroshima18} to model subhalo evolution, they take the value of $\mu(1)$ and $\eta(1)$ by assuming the subhalo initially has NFW profile with $\alpha=1$. In our model, the inner profile $\alpha$ of satellite before accretion is not fixed as $\alpha=1$, but varies with the star formation efficiency, as given from Eq.~\ref{eq:alpha}. To apply Eq.~\ref{eq:gx} on any $\alpha$, we use a linear interpolation to their best fitting values to get $\mu$ and $\eta$ for any $\alpha$ determined from Eq.~\ref{eq:alpha}. Note that in the simulation of \cite{Penarrubia10} the inner slope $\alpha$ is assumed to be fixed during the subhalo evolution, but as $V_{max}$ and $r_{max}$ are changed, one has to rescale its $\rho_{0}$ and $r_{0}$ in Eq.~\ref{eq:rhor} to solve the new $r_{max}$ and $V_{max}$ given by Eq.~\ref{eq:gx}. The procedure is given in the Appendix A of the \cite{Penarrubia10} paper. 

\begin{figure}
\centerline{\epsfig{figure=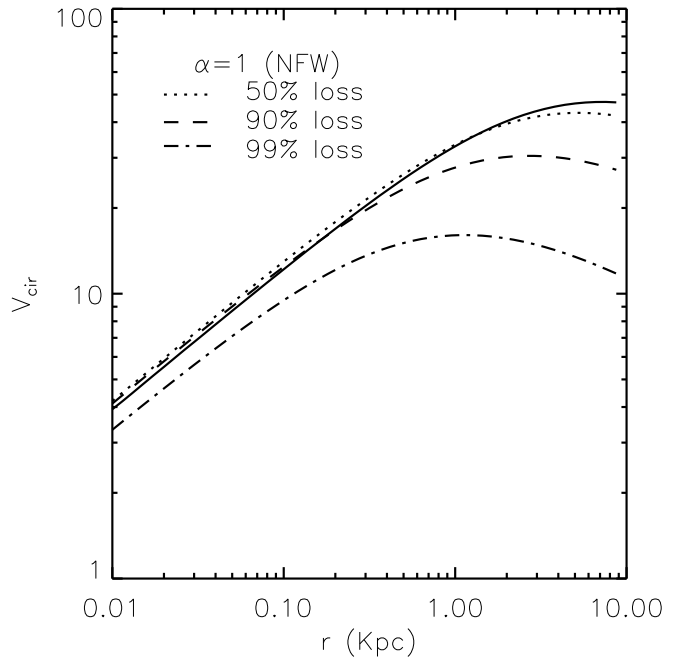,width=0.49\textwidth}}
\caption{The effect of tidal heating on the circular velocity of subhalo. Here predictions are based on the Pe{\~n}arrubia et al. (2010) simulations for a NFW halo with mass of $10^{10}M_{\odot}$ at redshift 1. Different lines are for different mass loss rate by tidal process.}
\label{fig:vcir_tidal}
\end{figure}
The above Eq.~\ref{eq:gx} depends on the mass loss of the satellite galaxy after accretion. Following \cite{Kang09}, we use the orbit-average mass-loss rate given by \cite{Giocoli08} to model the evolution of the subhalo mass, 
\begin{equation}
\frac {dm}{dt} = -\frac{m}{\tau}(m/M_{host})^{\zeta}
\label{eq:massloss}
\end{equation}
where $m$ and $M_{host}$ are the dark matter mass of the satellite and the host. We tune the parameter $\tau$ and $\zeta$ to best fit the subhalo mass function from simulations \citep[e.g.,][]{Gao04}. The free parameters ($\tau$,$\zeta$) in our model are ($3$ Gyr, 0.06), slightly different from those ($2$ Gyr, $0.07$) in \cite{Giocoli08}. This is due to the different merger tree algorithm used in our study. Using the above equation it is found that on average subhaloes have lost 90\% of their mass after accretion and the predicted mass-to-light ratio agrees well with the measured data \citep{Kang09}.

In Fig.~\ref{fig:vcir_tidal} we show the examples of tidal stripping on the circular velocity profile of a NFW halo with mass $10^{10}M_{\odot}$ accreted at redshift $z_{acc}= 1$. Different lines are for predictions with different fractions of mass loss. Note that here the fraction of mass loss is arbitrary set to show its effect on the circular velocity. It is seen that for subhalo with mass loss less than 90\%, the circular velocity is almost not affected within $1 Kpc$. In fact it is slightly higher in central region as the scale radius, $r_s$, is decreased with this fraction of mass loss. For subhalo losing 99\% of its mass, the circular velocity is strongly reduced at all radii. We will later see that the very faint satellites, such as Canes Venatici and Ursa Minor, have lost more than 90\% of their mass after accretion, so their circular velocity are strongly affected by the Milky Way tidal force. This is mainly due to the earlier accretion redshifts for faint satellites.

Now  with the  above formulae  at  hand, we  are able  to predict  the
circular velocities  of the  satellites after  their accretion.  We note
that the above model has some  limitations.  The first is that we do
not consider  the scatter on  the $c-M$  relation. The second  is that the fitting  formula, Eq.~\ref{eq:alpha}, for  baryonic effect  on the  halo inner  profile is  obtained only  from CDM  simulation, while in WDM model, the halo  concentration is lower and the  change of  inner profile  due to  baryonic feedback  could be
different. Nevertheless, \cite{Maccio19} recently have shown  that this effect
is very weak  in the WDM model,  so we still use  this fitting formula for the  WDM case. Third, the  exact tidal effect
on subhalo inner structure is more complicated and shoud depend on the
specific infall orbit, individual mass loss of each  satellite galaxy, but here we use the
average  mass loss  rate for  subhalo accreted  at given  time and  it
should be viewed as an average effect.

\section{Results}
\label{sec:results}
\begin{figure}
\centerline{\epsfig{figure=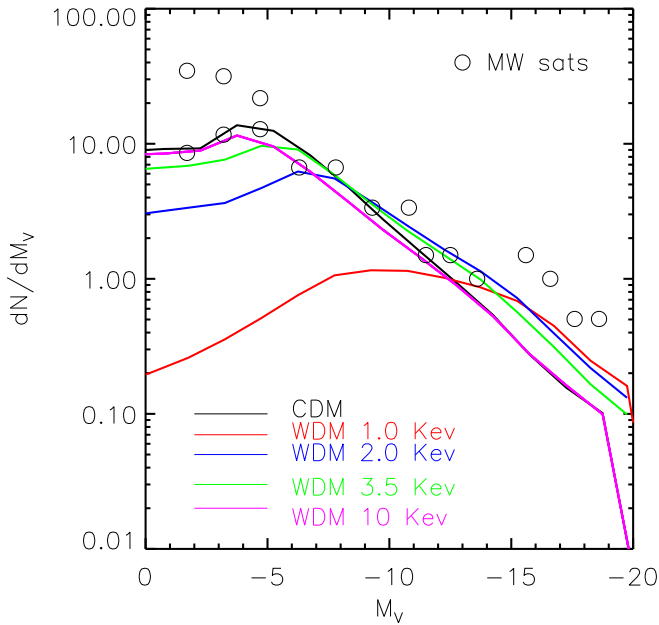,width=0.49\textwidth}}
\caption{The luminosity functions of satellite galaxies in MW type halo from the CDM and WDM models. The empty circiles are data from Koposov et al. (2008) and lines are our model predictions. For each model, we have tried to tune the parameters for star formation and feedback to best match the data. The line styles are the same as those in Fig.~\ref{fig:massfn_acc}. It is seen that $m_{s} = 1, 2$ keV models fail to match the MW data at the faint end.}
\label{fig:LFs}
\end{figure}

\subsection{satellite luminosity function}

\begin{figure*}
\centerline{\epsfig{figure=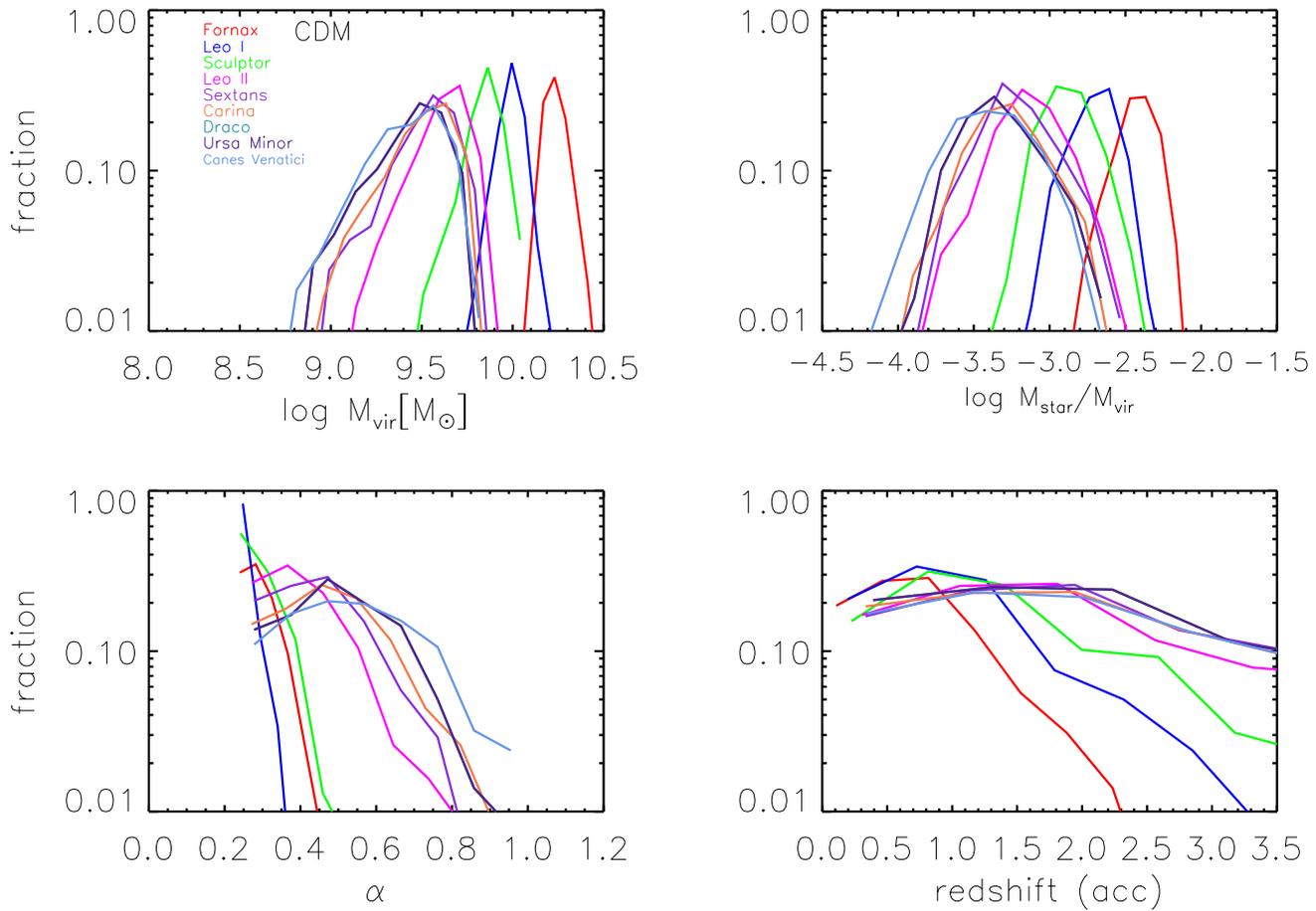,width=0.98\textwidth}}
\caption{The predicted properties of bright satellite galaxies from the SAM. For each observed bright satellite, we select its counterparts, which have the same luminosity as the observed one, from our model satellite galaxies in the 5000 Milky Way host haloes. The upper left, upper right, lower left panels are for the distributions of the halo mass, star formation efficiency, inner density profile at infall time. The lower right panel shows the accretion redshifts. Each color line corresponds to one satellite galaxy as labelled in the upper left panel.}
\label{fig:distribution}
\end{figure*}

In the semi-analytical model, the  free model parameters, such as star
formation and supernova feedback efficiency, are often tuned to match
the local observations, usually the stellar mass function and cold gas
content \citep[e.g.,][]{Kang05,Guo11,Luo16}. As
the observational constraints  on these free parameter  are very weak,
so they are coupled with the  cosmological model and the dark matter
properties. As shown in \cite{Kang13}, the observed stellar mass
function at $M_{\ast}>10^{9}M_{\odot}$ can also  be matched in the WDM
model with $m_{\nu}  > 1$ keV by tuning the  model parameters, and the
degeneracy can be broken by using  other data, such as the Tully-Fisher
relations. Here  we tune the supernova feedback efficiency to best match  the
satellite   luminosity  function in both the CDM and WDM models. It is found that we can get a good match to the bright end of the luminosity function by setting the feedback efficiency as 0.02, 0.05, 0.1 and 0.3 for the WDM model with $m_{\nu} = 1, 2, 3.5$ and $10$ keV. For the CDM model we set the feedback efficiency as 0.3.

In  Fig.~\ref{fig:LFs}  we show  the best  matched
satellite   luminosity  functions   from  the   CDM  and   WDM  models,
respectively. In agreement with previous results \citep[e.g.,][]{Kang09,Maccio10a}, it is possible for the CDM to match  the data up to $M_{v} = -5$. The data at $M_{v} > -5$ has larger uncertainty due to different assumptions on the intrinsic spatial distribution (NFW or isothermal) of satellite galaxies in the MW.  In this work we will only focus on brighter satellites with $M_{v} < -5$. It is  seen that  for the WDM model with $m_{\nu} \sim 1-10$ keV,  the
bright  end of  the  luminosity  functions can  be roughly reproduced. In  the
semi-analytical  model, galaxy stellar  mass is  very sensitive  to the
supernova feedback efficiency.  As the formation of low-mass haloes in
WDM model  is suppressed, to form  enough stars  in low-mass
haloes, we  need to  lower the feedback efficiency  in those
haloes.  For example,  in the  WDM model  with $m_{\nu}  =  1$ keV, the
supernova feedback efficiency is decreased  to about 2\%. Note that the feedback efficiency is highly uncertain, ranging between $0.1$ to  $0.5$ adopted in different SAMs \citep[e.g.,][]{Somerville99,Cole00,Kang05,Croton06,Guo11} and simulations \citep[e.g.,][]{Limiao17}. From  Fig.~\ref{fig:LFs} it is seen that  even with such a lower
feedback, the WDM model with $m_{\nu} =1$ keV  still fails to  reproduce  the  number of  very  faint
satellites.  The same behavios is also seen for the $m_{\nu} = 2$ keV model. The WDM model with $m_{\nu} \geq 3.5$ keV is able to reproduce enough faint satellite galaxies.

\subsection{The host haloes of satellite galaxies}

\begin{figure*}
\centerline{\epsfig{figure=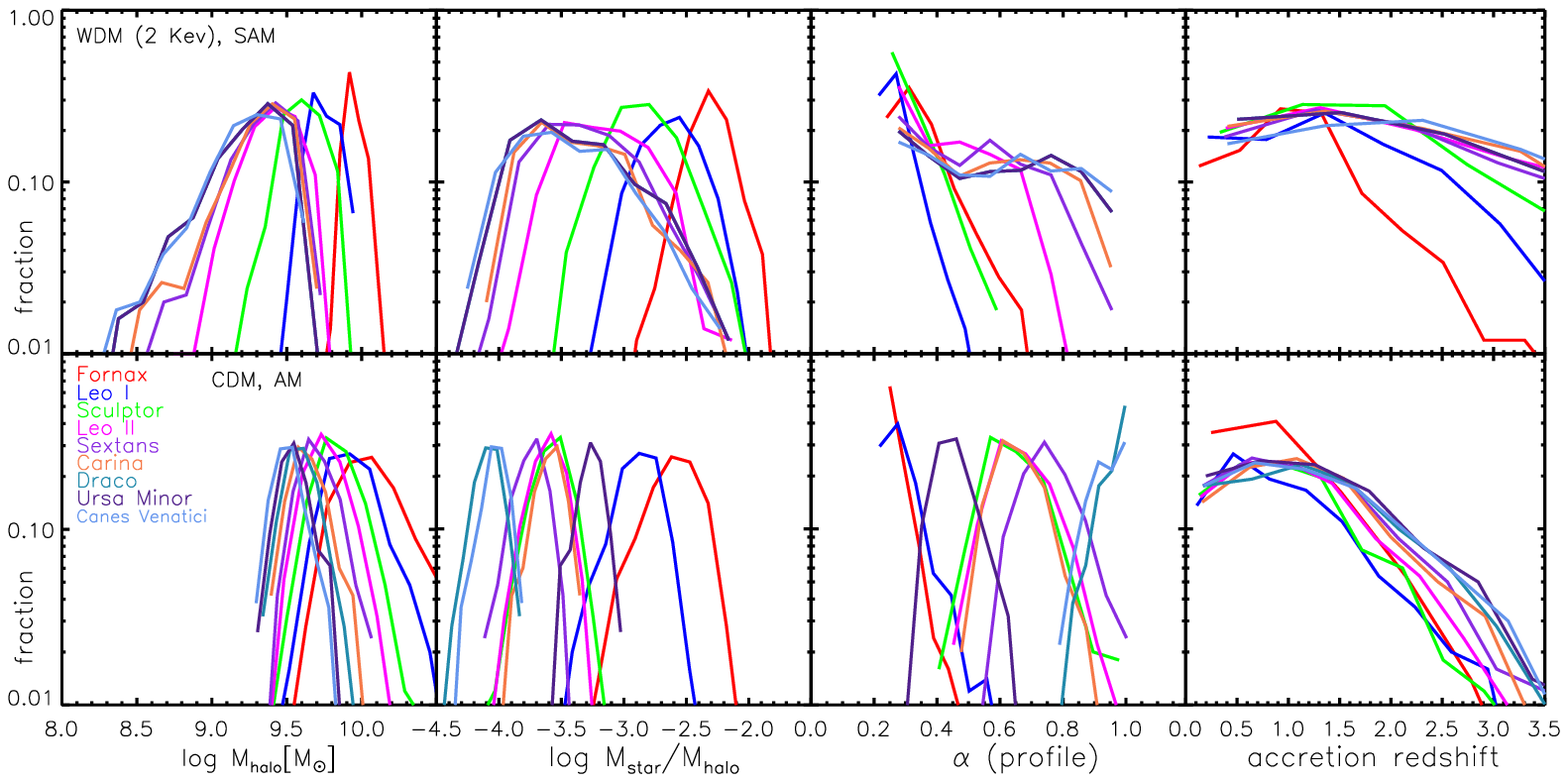,width=0.98\textwidth}}
\caption{The same as in Fig,~\ref{fig:distribution}, but for the WDM model with $m_{s}=2$ keV (upper panel). The lower panel is for CDM but where the satellites are selected using the abundance matching method so that the brightest satellites (with their observed stellar mass) are assigned to the most massive subhaloes at accretion.}

\label{fig:distribution1}
\end{figure*}

Using our model we are also able to predict the other properties of the MW satellites. For each observed satellite galaxy, we select its counterpart from each of the 5000 MW host galaxies. Here we define the model counterparts as those satellites having luminosity as the observed one. Note that the realistic counterparts of an observed satellite galaxy in the model should be those having the same luminosity and phase-space distribution (spatial position and motion) as the observed one. Since our model does not include the phase-space information, here we only use the luminosity to select model satellites. In Fig.~\ref{fig:distribution} we show the distributions of the halo mass, star formation efficiency, and halo inner density slope at the infall time for model satellites correspnding to each of the nine classic satellites. The lower right panel shows the distributions of the accretion redshift. As in previous work, we have exclude the Large Magellanic Cloud and Small Magellanic Cloud from our analysis since the two massive satellites are rare in MW type haloes \citep[e.g.,][]{Liu11,Jiang12}. We also exclude Sagittarius galaxy as it is now in process of being disrupted. Here the model predictions are from CDM and satellites are ordered in decreasing luminosity with different colors, as labelled in the upper left panel.

It is clearly seen from upper left panel of Fig.~\ref{fig:distribution} that in the SAM, more luminous satellites are on average living in more massive subhaloes. Most faint satellites, such as Leo II and fainter ones, stay in haloes with infall mass around $3\times 10^{9}M_{\odot}$.These predicted halo mass at infall agree well with those derived from hydro-simulation \citep{Buck19}. The upper right and lower left panels show the star formation efficiency and associated inner DM density profiles. It is found that the star formation efficiency is also higher in the luminous satellites and the creation of shallow density profiles are stronger in these galaxies. It is interesting to see that the inner profile of satellite Leo I is strongly affected by the baryonic effect, as its star formation effieicncy $M_{\ast}/M_{halo}$ is around $0.002$, close to the peak shown in Fig.~\ref{fig:fittingformula}. 

In Fig.~\ref{fig:distribution1} we show the same predictions from the SAM in the WDM model with $m_{\nu} = 2$ keV and from the halo AM method in CDM model, in the upper and lower panels respectively. By comparing the semi-analytical predictions from the  CDM and WDM models (Fig.~\ref{fig:distribution} and upper panels in Fig.~\ref{fig:distribution1}), it is found that the host halo mass of satellites from WDM is systematically lower by $0.2 dex$ than that from the CDM model. This is not surprising, as in the WDM model the number of haloes is lower, so for given number density of subhalo, the average subhalo mass is lower by $0.1 \sim 0.3 dex$ in the WDM $2.0$ keV model, as can be seen from Fig.~\ref{fig:massfn_acc}. Meanwhile, the host halo mass distribution is slightly wider in the WDM model. The decreases in host halo mass leads to a slightly higher and wider distribution of the star formation efficiency in the WDM model (second panel). For some fainter satellites, the star formation efficiency in WDM model is shifted to higher value than the peak, leading to a slightly steeper slope $\alpha$. 

By comparing the lower right panel of Fig.~\ref{fig:distribution} with the upper right panel of Fig.~\ref{fig:distribution1}, it is seen that in the WDM model, the accretion redshifts for satellite galaxies are slightly higher, especially for the brightest ones. This is mainly due to the lower host halo mass in the WDM and it is well know that in either WDM or CDM, low-mass haloes form earlier. Overall, the host halo mass and accretion time of satellites from the WDM 2.0 keV model and the CDM are not significantly different, and we will later see that it is mainly the difference in the halo concentration leading to the low circular velocity in the WDM model. 

The lower panels in Fig.~\ref{fig:distribution1} show the predictions from the halo AM in the CDM model. As noted before, we assign the measured stellar mass of the nine classic satellites (see Misgeld \& Hilker 2011 and references therein) to the most massive nine subhaloes in the model. Note that we have omitted the two most massive subhaloes in each host halo from our model, as they are thought to hold the LMC, SMC. It is seen that the distributions are significantly different from those in Fig.~\ref{fig:distribution}. Firstly, the host halo mass of satellites in the AM model are on average larger than those in the SAM, and the mass distribution becomes narrower. This is not surprising from the AM method. For most of the faint satellites, the star formation efficiency in the AM model is lower, mainly due to the higher host halo mass. Secondly, the inner density profile of satellite ($\alpha$) is now on average more close to NFW, as the star formation efficiency and the associated feedback is now lower in the AM model. Thirdly, the accretion redshifts of satellites are lower, with most are accreted below redshift $3.5$, while in the SAM some satellites are accreted at much earlier times. One consequence of the later accretion redshift means the effect of tidal stripping, both dependent on the subhalo mass and inner mass distribution, is weaker in the AM model. Thus we will later see that the decrease of circular velocity of satellites by baryonic feedback or tidal heating in the AM model is weaker than that in the SAM. 

We note that some of the above results, such as satellites infall time, are not comparable to those derived from the real data. One important caveat is that we do not have the phase-space information of the satellites, but only the stellar mass of the satellites. \cite{Rocha12} have tried to derive the infall times for MW satellites based on their dynamical properties. They found that Carina, Ursa Minor and Sculptor were accreted more than 8 Gyr ago ($z > 1$). Fornax is recently accreted around $\sim 2$ Gyr ago. The remaining othere satellites, including Sextans and Segue 1, are probably accreted early, but with larger uncertainty. Using Gaia data the infall times and orbits for most satellites have been recently updated \citep{Fritz18,Fillingham19}. Satellites in our SAM have on average higher accretion redshifts than the derived ones, but the genetal trend is qualitatively consistent with the data in the sense that faint satellites, such as Carina and Ursa Minor, have higher accretion redshifts than those luminous satellites. This is the consequence of two combined effects. First, faint satellites are more likely to form in low-mass haloes which were accreted more early in the CDM model. Second, in case of cosmic reionization, faint satellites forms more earlier as their host halo is capable of holding hot gas at earlier times.

\subsection{satellite kinematics}

\begin{figure*}
\centerline{\epsfig{figure=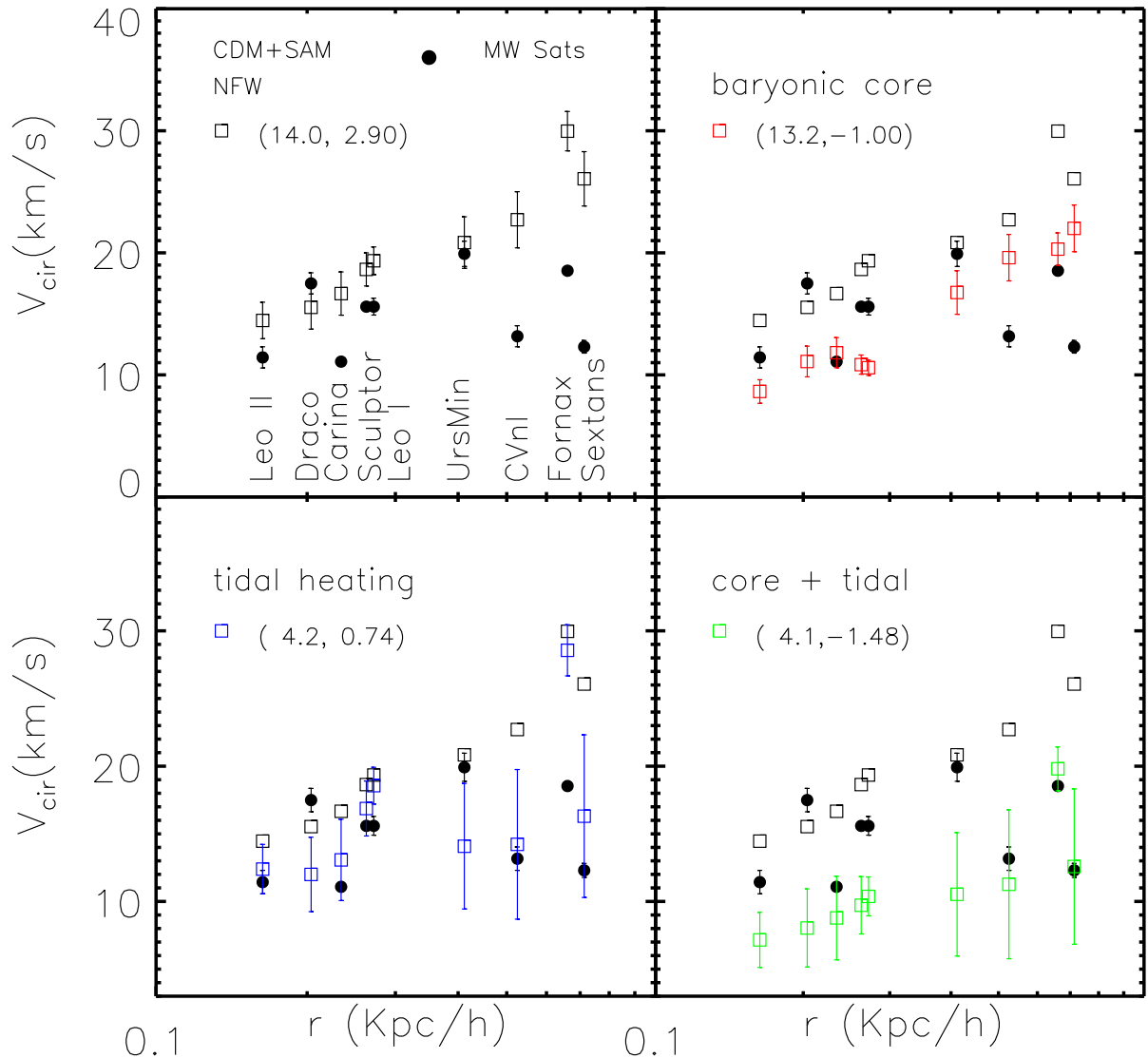,width=0.9\textwidth}}
\caption{The circular velocity of satellite galaxies at the half-light radii. The black dots are the observational data. The squares in each panel are predictions from the semi-analytical model in the CDM model and the error bars are the velocity spread obtained from Eq.~\ref{eq:spread}. For the black squares (replicated in all panels), satellite galaxies have NFW profile and no baryonic feedback or tidal heating is included. The color squares in upper right and lower left show the results with only baryonic feedback or tidal heating effect included, respectively. The green squares in lower right have included both effects. In each plot, the degree of agreement between data and model, $(\chi_{red}^{2}, \delta)$, is also labelled.}
\label{fig:rvc-CDM-sam}
\end{figure*}

In this section  we compare the  predicted kinematic  properties of  the
satellites, namely the circular  velocity, to the data.  Observational
work  \citep[e.g.,][]{Mateo98,Walker09,Wolf10,McConnachie12} have  measured  the half-light  radii and  velocity
dispersions  for  most satellite  galaxies  in  the MW.  Following  the
arguments of \cite{Boylan-kolchin11}, we focus on  the data for bright satellite  galaxies with  $L_{V} > 10^{5}L_{\odot}$  which have
more   reliable   measurements   of   stellar   spectra   and   member
identification.   For   each  satellite   we  transfer   the  measured
line-of-sight velocity dispersion to the circular velocity at the half
radii ($r_{1/2}$), labeled  as $V_{o} = \sqrt{3}\sigma_{los}$.   The data of
$r_{1/2}$ and  $\sigma_{los}$ for  the nine  bright satellites  can be
found from the paper by \cite{Wolf10}. We note that this mass estimate may suffer from a systematic bias, as is found in \cite{Campbell17}.

For each model satellite, we calculate the circular velocity at the half light raii, labelled as $V_{m}$, using the above prescriptions on the evolution of the density profile. As seen from previous section, the model counterparts of each observed satellites have a wide distribution of properties, such as infall mass, infall time and tidal stripped mass, so the predicted $V_{m}$ will also have a range of values. To compare with the data, we get the median velocity, $V_{m,50}$, defined as the circular velocity at 50 percentage of the velocity distribution. We also obtained the spread of the velocity as, 
\begin{equation}
\sigma_{m}= (V_{m,90}-V_{m,10})/2
\label{eq:spread}
\end{equation}
where  $V_{m,90}$,  $V_{m,10}$  is  the  circular  velocity  of  model
satellites at the 90 and  10 percentage of the distribution. To quantify the degree  of agreement between
the model  prediction and the  data, we adopt the reduced chi-squared \citep{Andrae10} as,
\begin{equation}
\chi_{red}^{2}=\frac{1} {n}\Sigma\frac {(V_{m,50}-V_{o})^{2}} {\sigma_{o}^{2}+\sigma_{m}^{2}}
\end{equation}
where $\sigma_{o}$ is the uncertainty of the measured circular velocity of observed satellite, $n$ is  the total number of observed satellite
galaxies. From the distribution of $\chi_{red}^{2}$ and for 9 degree of freedom (assuming the data of nine satellites are independent), the chance of a model matching the data with probability of 5\%, 1\% and 0.1\% corresponds to $\chi_{red}^{2} = 1.87, 2.4$ and $3.2$. In this paper we select $\chi_{red}^{2} = 3.2$ as a upper limit to exclude or accept a model. We note that given the systematics in our study, the choice of statistical significance should be taken as suggestive only. However, as $\chi_{red}^{2}$ does not fully specify the agreement between the data and the model, here we define a  mean deviation $\delta$ as,
\begin{equation}
\delta = \frac{1} {n}\Sigma\frac {(V_{m,50}-V_{o})} {(\sigma_{o}^{2}+\sigma_{m}^{2})^{0.5}}
\end{equation} 
The defined $\delta$ quantifies the systematic deviation between the
model  and the  data, in  which $\delta  > 0$  indicates  the predicted
circular velocity is systematically higher than the data.  To place a rough limit on the agreement between model and data, we set $\chi_{red}^{2} = 3.2$ and $-1 < \delta < 1$ as the threshold to exclude an model. In each following plot we label the values $(\chi_{red}^{2}, \delta)$ as a reference of how the model agrees with the data.

\begin{figure*}
\centerline{\epsfig{figure=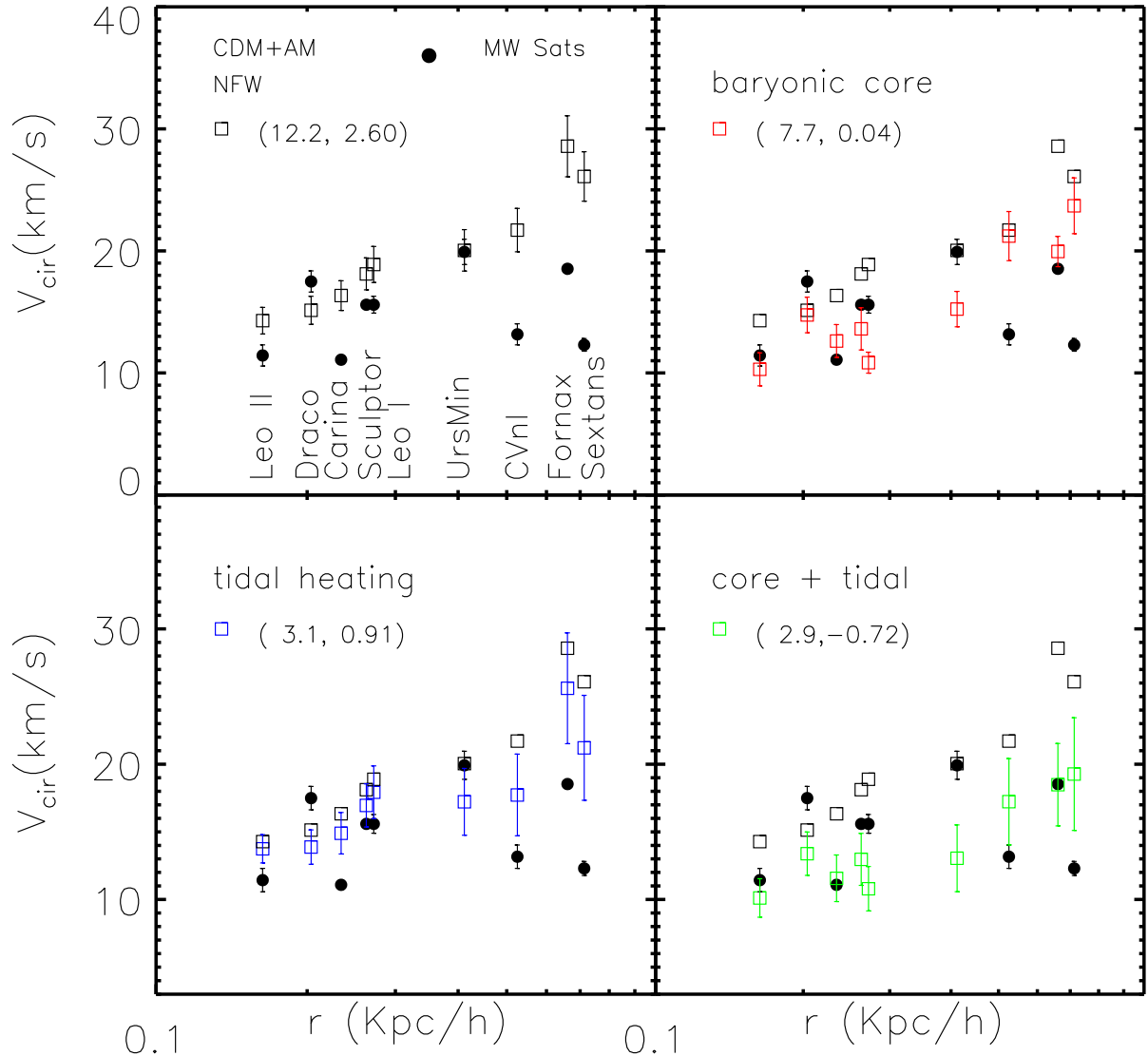,width=0.9\textwidth}}
\caption{As in Fig.~\ref{fig:rvc-CDM-sam} for the CDM mode, but here the satellite galaxies are selected based on their halo mass at accretion. Compared to the predictions from the SAM in Fig.~\ref{fig:rvc-CDM-sam}, it is seen that the baryonic feedback and tidal heating effects are both weaker. This is because satellites are now living in more massive subhaloes, thus both the star formation efficiency and accretion redshifts are also lower.}
\label{fig:rvc-CDM-macc}
\end{figure*}

Now we compare the mode predictions with the data. We show the circular velocity of the nine classic satellites at their half-light radii in Fig.~\ref{fig:rvc-CDM-sam}, where we separate the effect of baryonic feedback and tidal heating in different panels. The black circles are the data for observed satellites \citep[e.g.,][]{Wolf10}. The squares in each panel are the model predictions from our SAM in the CDM model with error bars showing the velocity spread obtained from Eq.~\ref{eq:spread}. The black squares in the upper left panel are for satellites with NFW density profiles, with neither baryonic feedback nor tidal heating included, and they are replicated in other panels (no errorbar). In the upper right panel only baryonic feedback, or equally the new profile given from Eq.~\ref{eq:alpha}, is included. In the lower left panel, only the effect of tidal stripping and heating from Eq.~\ref{eq:gx} is included. Note that here no bayonic feedback included, so each subhalo has a NFW profile ($\alpha=1$). In the lower right panel, both effects are included. In this case, we firstly consider the effect of baryonic feedback, obtaining a star-formation dependent $\alpha$, and then include the tidal heating. This is reasonable as star formation in satellites is often strongly suppressed after infall. As shown in \cite{Penarrubia10}, the tidal effect on satellites inner structure depends on the initial profile $\alpha$, with more strong effect for lower $\alpha$, thus the predictions in the lower right panel is not the sum of the two effects in the upper right and lower left panel, but being slightly stronger.

Fig.~\ref{fig:rvc-CDM-sam} shows that in the CDM model the predicted circular velocities of model satellites with neither baryonic feedback  nor tidal heating are systematically  higher  than observed (upper left panel).  In particular, the disagreement is more server for the satellite Carina, CVnI, Fornax and Sextans. This was  already clearly  shown  by  many  previous studies  \citep[e.g.,][]{Boylan-kolchin11,Wetzel16}. In fact these four satellite galaxies are the main drivers of the TBTF problem, as they contribute mostly to the relative flat distribution of the circular velocities.  With only baryonic feedback included (red squares in the upper right panel), it is seen that the feedback does reduce the circular velocities by creating shallow profiles in some satellite galaxies. The degree of decrease varies among the satellites. For example, it is prominent in Leo I, but weak in UrsMin and CVnI. This is due to the relation between density profile $\alpha$ and the star formation efficiency in the satellites, as can be seen from Fig.~\ref{fig:distribution}. For some satellites, such as Carina and Fornax, the baryonic feedback leads to better agreement between the model and the data. However, for other satellites, the baryonic feedback is too strong, leading to the circular velocity being lower than the data. The values of the statistical measures we set to compare with the data ($\chi_{red}^{2} = 13.2,\delta = -1$) shows that the agreement between data and model is still not very satisfying.

By only including the tidal heating (blue squares in lower left panel), it is seen that the velocity distribution is now much flatter, in better agreement with the data, as also indicated by the statistical measures ($\chi_{red}^{2} = 4.2$, $\delta=0.74$). In particular the circular velocities of Carina, CVnI and Sextans are strongly decreased to match the data. This seems to suggest that tidal heating is the main factor to solve the high circular velocity of the main drivers of the TBTF problem. In this case, we should expect to see clear tidal feature in these satellite galaxies. However, observational determination of tidal feature in satellites is difficult and studies \citep[e.g.,][]{Lokas12} have shown that some satellites, such as Carina, have experienced strong tidal stirring. \cite{Roderick16} found that tidal disruption is not prominent for Sextans, but tidal stirring effect could still be present. We will briefly discuss this in the final section. Note that the error bar in presence of tidal heating  is larger than those from the baryonic feedback. This is because the distribution of satellite accretion time is wider, leading to a wider range of mass stripping and associated tidal heating. More interesting results are seen in the lower right panel where both baryonic feedback and tidal heating are included. Now the predicted satellite velocity is systematically lower than the data, as indicated by the value $\delta = -1.48$. Apparently, the problem is now reversed, from TBTF to too-diffuse-to-pass, as mentioned in previous studies \citep{Dutton16}. This is mainly due to the strong feedback effect in this model, and we will discuss more in the last Section. 

Now we investigate the results from the AM method in the CDM model, shown in Fig.~\ref{fig:rvc-CDM-macc}. As previously shown, the main change in the AM method is that on average the host subhalo mass is larger than those in the SAM. The higher subhalo mass leads to two main consequences. First, the accretion redshifts of subhaloes are lower, thus the tidal stripping effect, depending on time after infall, is weaker in this model. Second, the star formation efficiency is now lower, thus the halo inner density profile $\alpha$ is more close to NFW.  The upper left panel shows that in this model, the circular velocity distribution is very similar to that in the SAM. This is not surprising, as we have shown in Fig.~\ref{fig:cmrelation} that in the CDM model the $c-M$ relation in the CDM model is relatively flat. 

\begin{figure*}
\centerline{\epsfig{figure=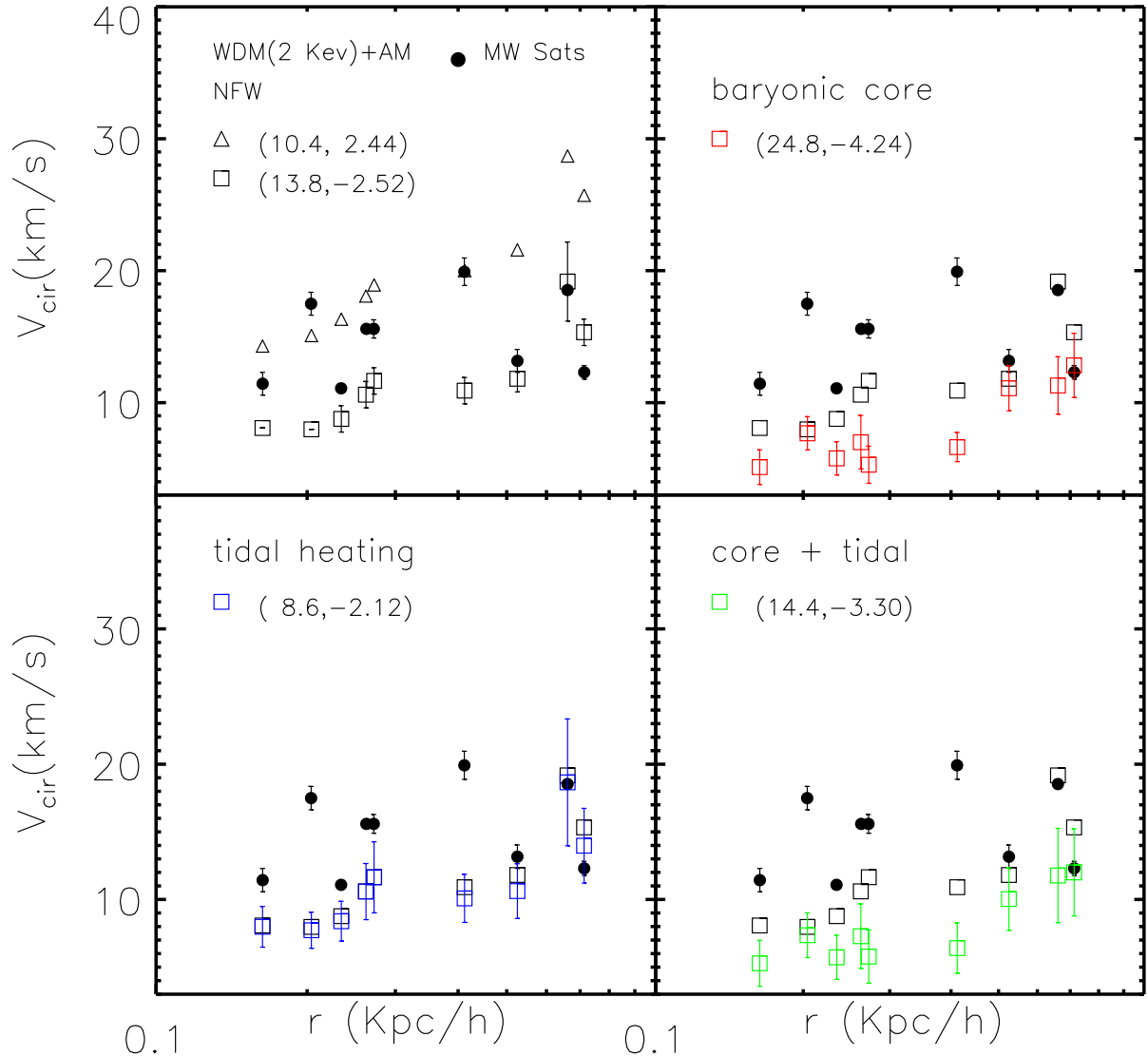,width=0.9\textwidth}}
\caption{As in Fig.~\ref{fig:rvc-CDM-macc}, but for the WDM model with $m_{\nu} = 2.0$ keV. Here in the upper left panel, we show additional predictions using the triangles in which the $c-M$ relation is the same as the CDM. So the comparison between triangles and squares directly shows the effect of halo concentration on circular velocity. The value of ($\chi_{red}^{2},\delta$) in each panel indicates that the model predictions are systematically lower than the data.} 
\label{fig:rvc-WDM-macc-2keV}
\end{figure*}
The upper right panel in Fig.~\ref{fig:rvc-CDM-macc} shows that with baryonic feedback, some satellites will have lower velocities than the data, such as the Sculptor and Leo I. Overall, the baryonic effect in this AM method is weaker than that in the SAM, and the fitted value ($\chi_{red}^{2}=7.7$, $\delta=0.04$) indicates that the model is still inconsistent with the data. With careful inspection, we find that for the satellites mainly responsible for the TBTF problem in the MW, such as CVnI and Sextans, the baryonic effect is negligible and the model predictions for them are still away from the data. However, with tidal heating effect included, as shown in the lower left panel, it is seen that the model predictions agree slightly better with the data, either for each single satellite or the overall distribution. With both baryonic feedback and tidal heating included, the velocity distribution shown in the lower right panel is more flat, more consistent with the data.

Overall, our above results show that in the CDM model, the effects of baryonic feedback and tidal heating depend on the model for galaxy formation. In the SAM, the host subhaloes of satellites have wide distributions in mass and accretion redshifts, giving rise to stronger baryonic feedback and tidal heating effects. In this case, the predicted circular velocity of satellite galaxies are systematically lower than the data. In the AM model, satellites form in most massive subhaloes, so the effects of baryonic feedback and tidal heating are modest. 
However, in both cases, it is found that tidal heating effect must be invoked to lower the circular velocities of some satellites, such as CVnI and Sextans, which are thought to be the main drivers of the TBTF problem.

\subsection{Satellites kinematics with warm dark matter}

\begin{figure*}
\centerline{\epsfig{figure=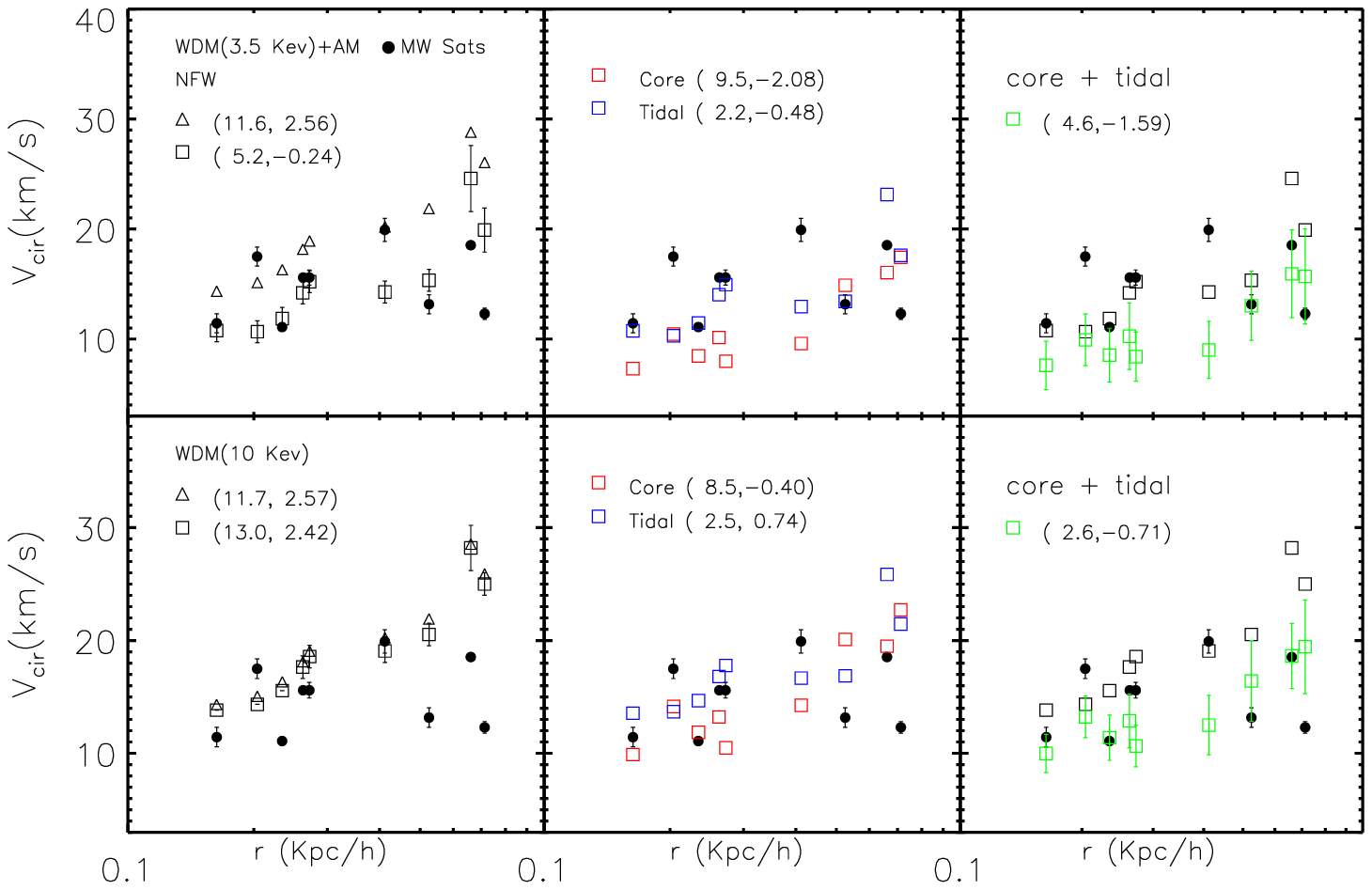,width=0.9\textwidth}}
\caption{As in Fig.~\ref{fig:rvc-WDM-macc-2keV}, but for two other WDM models with $m_{\nu} = 3.5, 10$ keV. It is seen that if both baryonic feedback and tidal heating are ineffective, the 3.5 keV WDM model is marginally acceptable. If both effects are in act, the 3.5 keV model can be excluded (for MW halo mass of $10^{12}M_{\odot}$), and the 10 keV model can not be excluded.}
\label{fig:rvc-WDM-macc}

\end{figure*}
As mentioned in the introduction, WDM model is proposed as an alternative solution to the TBTF problem. However, most existed studies using WDM model often neglected the effect of baryonic feedback and tidal effect, but see recent studies by \cite{Bozek19} and \cite{Maccio19}. The analysis made in the previous section has shown that the baryonic feedback and tidal effects depend on how we selected satellites. In the AM method those effects are minimized as satellites are from the massive subhaloes which on average have lower accretion redshifts and weak baryonic feedback effect (due to their high halo mass). Thus, the AM case is a reasonable benchmark that could provide a conservative (in terms of the impact of baryonic feedback) lower limit in the WDM mass. Thus in this section, we use the AM method and apply it to the WDM model to set a lower limit on the WDM mass. Here we only consider three cases with $m_{\nu} = 2, 3.5, 10$ keV. We do not show the results for $m_{\nu} = 1$ keV as it can be ruled out just from the satellite luminosity function shown in previous section. Also note that in this part we fix the host halo mass as $10^{12}M_{\odot}$ and we will discuss the implications of other host halo mass in section~\ref{sec:hostmass}.

We firstly show the results from the WDM model with $m_{\nu} = 2$ keV in Fig.~\ref{fig:rvc-WDM-macc-2keV}. Similar to Fig.~\ref{fig:rvc-CDM-macc}, we isolate the baryonic feedback and tidal heating in different panels. In the upper left panel where the dark matter density profile is modelled as NFW, we also add additional predictions using the black triangles, in which we use the $c-M$ relation from the CDM model. This is purely for comparison between the CDM and WDM to see the effects of changing halo concentration. It is found that the triangles are very similar to the squares in Fig.~\ref{fig:rvc-CDM-macc}, indicating that although the halo formation history is different in the WDM model, the circular velocity from NFW profile is very similar. This is mainly due to the flat $c-M$ relation in the CDM model. The squares show that the circular velocities of satellites can be greatly reduced due to lower concentration in the WDM case. However, the statistical measures ($\chi_{red}^{2}=13.8$, $\delta=-2.52$) in our results shows that the model predictions are still systematically lower than the data. 

The upper left and lower right panels show the effects of baryonic feedback and tidal heating, respectively. It is found that the baryonic feedback is more strong than the tidal heating, resulting in satellite circular velocities too below the data. The lower right panel includes both effects. Apparently, this plot shows that the WDM model with $m_{\nu} = 2$ keV can be safely ruled out. It indicates that in the $2$ keV WDM model, any baryonic feedback effect and tidal heating is not tolerated, which is apparently implausible in reality.

We further show the results for two other WDM models with $m_{\nu} = 3.5$, $10$ keV in Fig.~\ref{fig:rvc-WDM-macc}.For the 3.5 keV model one can find the upper middle panel that with only tidal effect, the agreement between the model and the data is acceptable ($\chi_{red}^{2}=2.2$, $\delta=-0.48$). With both feedback and tidal included, the model predictions are systematically lower than the data, seen from the upper right panel. For WDM model with $m_{\nu} = 10$ keV, the model can match the data with inclusion of both baryonic feedback and tidal process, as seen from the lower right panel. This plot shows that for host halo mass of $M_{vir} = 10^{12}M_{\odot}$ the WDM model with $m_{\nu} \leq 3.5$ keV can be excluded if baryonic feedback and tidal heating are effective. Our model can not exclude the model with $m_{\nu} = 10$ keV. 


\begin{figure*}
\centerline{\epsfig{figure=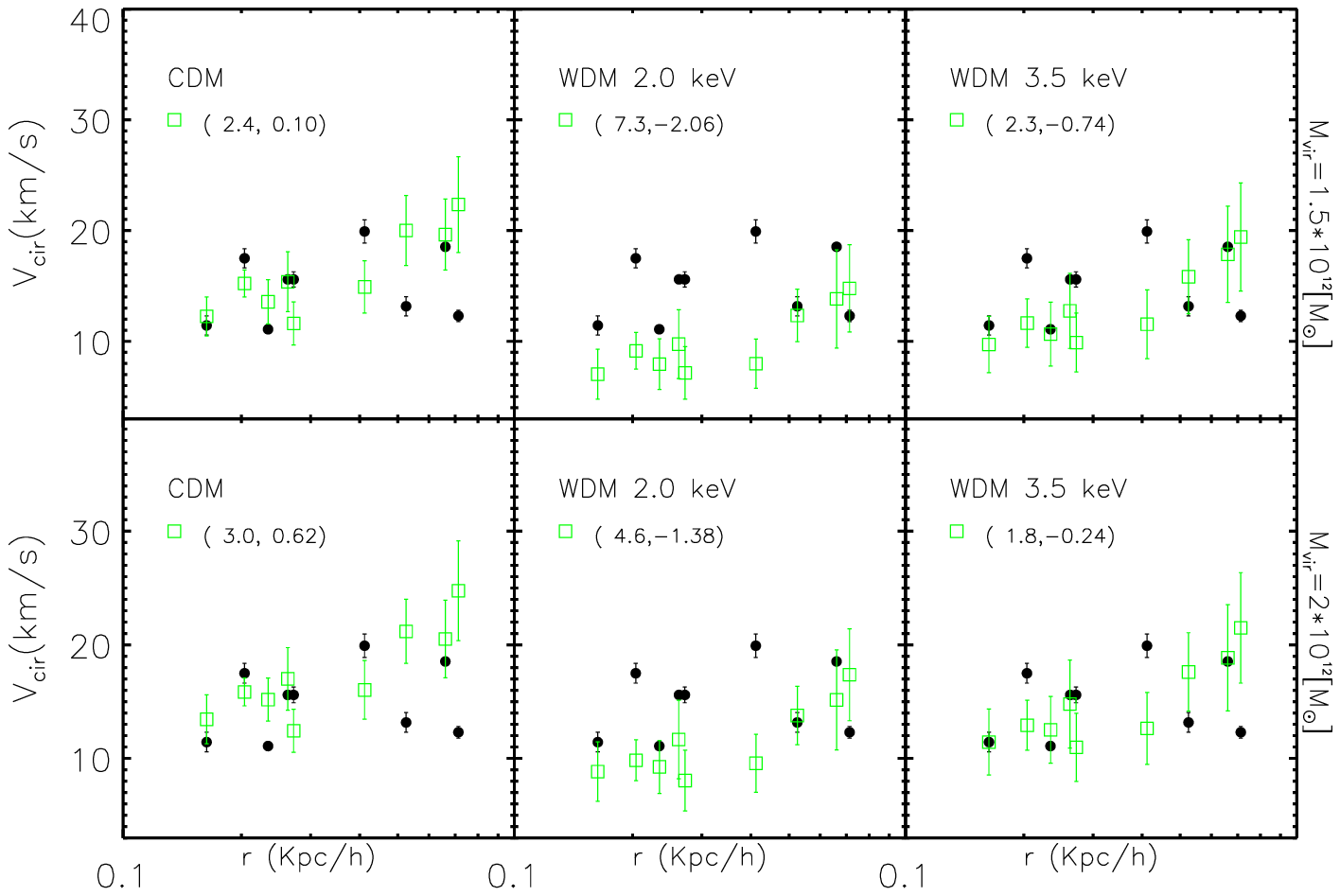,width=0.9\textwidth}}
\caption{The effects of higher host halo mass on satellite circular velocities. Here both baryonic feedback and tidal heating are included. The upper and lower panels are for two MW halo mass with $1.5\times 10^{12}M_{\odot}$ and $2\times 10^{12}M_{\odot}$ respectively. For the CDM model, a lower MW halo mass is favoured. The WDM model with 2 keV mass can be excluded in both cases and for WDM with 3.5 keV, it requires the MW halo mass to be larger than $1.5\times 10^{12}M_{\odot}$.} 
\label{fig:rvc-CWDM-macc-halomass}
\end{figure*}

\subsection{The effects of host halo mass}
\label{sec:hostmass}
For all results above we assume the MW has a halo mass of $10^{12}M_{\odot}$. Although most observational constraints on the MW mass are around this value, there is still uncertainty of a factor of 2 \citep{Li17,Callingham19}. For the WDM model with mass around a few keV, it is found from the satellite number count that the MW should have a mass larger than $10^{12}M_{\odot}$ \citep[e.g.,][]{Kennedy14}. Here we test if the constraints from satellite kinematics in WDM model are consistent with previous conclusions.

In Fig.~\ref{fig:rvc-CWDM-macc-halomass} we show the results for two MW mass, one is $1.5\times 10^{12}M_{\odot}$ in the upper panels and one is $2\times 10^{12}M_{\odot}$ in the lower panels. We do not test a MW halo mass lower than $10^{12}M_{\odot}$ as it is seen from previous results that with a halo mass of $10^{12}M_{\odot}$, the CDM predictions is slightly lower than the data even in the AM method. The left, middle and right panels are for CDM, WDM with $m_{\nu} = 2.0, 3.5$ keV respectively. Here we do not show the 10 keV model as it predictions are close to the CDM results, and we only plot the model predictions with both baryonic feedback and tidal heating included.

We firstly investigate the effects of host halo mass in the CDM. By comparing the lower right panel of Fig.~\ref{fig:rvc-CDM-macc} with the left panels in Fig.~\ref{fig:rvc-CWDM-macc-halomass}, it is seen that MW with a slightly larger mass around $1.5\times 10^{12}M_{\odot}$ agrees better with the data, with smaller $\chi_{red}^{2}$ and a $\delta$ close to 0. However, a more massive MW with mass of $2\times 10^{12}M_{\odot}$ gives slighter higher circular velocity for satellites, beginning to depart from the data. It shows that, with inclusion of baryonic feedback and tidal heating, a MW with mass around $1.5\times 10^{12}M_{\odot}$ agrees better with the MW satellite kinematics in the CDM model. 

For the WDM models as shown in the middle and right panels, it is seen that for WDM with $m_{\nu} = 2$ keV, a larger MW halo with mass of $2\times 10^{12}M_{\odot}$ still predicts circular velocities systematically lower than the data. Using satellite number count as a constrain, \cite{Kennedy14} found that the lower limit of MW mass is $1.7\times 10^{12}M_{\odot}$. Our results show that the mass limit can be pushed to a higher value. For WDM model with $m_{\nu} = 3.5$ keV, the right panels show that a MW with mass of $1.5\times 10^{12}M_{\odot}$ is acceptable, while a higher mass of $2\times 10^{12}M_{\odot}$ gives slightly better agreement with the data. We do not test higher values of the MW mass, as $2\times 10^{12}M_{\odot}$ is almost the upper limit from most observational constraints. Also the work of \cite{Kennedy14} shows that a higher MW mass larger than $2\times 10^{12}M_{\odot}$ will produce more bright satellites than observed.

\section{Conclusion and Discussion}
\label{sec:discuss}

In this paper, we present an analytical model to study the luminosity and kinematic properties of the Milky Way satellite galaxies.  We use a Monte-Carlo method to produce formation history for a large sample of Milky Way type haloes with mass around $10^{12}M_{\odot}$, and use analytical models to assign stellar mass to the subhaloes. Our model includes two key process to describe the evolution of the satellite density profile. One is the baryonic feedback which can induce a shallow density profile depending on the star formation efficiency, as suggested by recent state-of-the-art hydrodynamical simulations. The other is the tidal stripping process which will gradually strip the dark matter mass of the satellite galaxy and the associated tidal heating will redistribute the inner mass of the satellite. With these descriptions, one is able to predict the circular velocity of the satellites and compare the model with the data.  By applying the analytical model to both CDM and WDM models with different mass, we have obtained the following main results, 

\begin{itemize}
\item In the CDM model, galaxy formation model including cosmic re-ionization effect to suppress the baryonic content in low-mass haloes can fit the luminosity function of the satellite galaxies in the Milky Way. By tuning the model parameters for star formation and feedback, the WDM models with $m_{\nu} > 2$ keV can reproduce the luminosity distribution of bright satellites ($M_{V} < -5$), but a lower mass with $m_{\nu} \leq 2$ keV fails to produce enough faint satellites and can be excluded.


\item In the CDM model, by assuming that satellite galaxy initially follows NFW profile and neglecting the baryonic feedback and tidal process, the predicted circular velocities of the bright satellites are systematically higher than the data, in agreement with previous studies. The main drivers of this discrepancy between data and model (or the TBTF problem) are Carina, CVnI, Fornax and Sextans. By invoking only baryonic feedback in our model (semi-analytical or halo abundance matching), some satellites are predicted to have circular velocities, which are too low to be consistent with the data, while some satellites, such as CVnI and Sextans, always have high circular velocity than the data. It is found that tidal heating must be efficient to reduce the circular velocities of the main driver of TBTF, so as to agree with the data. The effect of tidal heating is in agreement with the findings from hydro-dynamical simulations \citep[e.g.,][]{Garrison-Kimmel19}.


\item In agreement with previous results \citep[e.g.,][]{Kennedy14} based on satellite number count, the constraint on WDM mass depends on the host halo mass. For MW with halo mass around $10^{12}M_{\odot}$, the WDM model with $m_{\nu} = 2$ keV can be excluded as the circular velocity of satellite galaxies are systematically lower than the data, even without any baryonic feedback or tidal heating. This is mainly because the halo concentration in this WDM case is too low. For model with $m_{\nu} = 3.5$ keV, the model prediction marginally agrees with the data if baryonic feedback and tidal heating are both ineffective. If both effects are in play, the MW halo mass should be larger than $1.5\times 10^{12}M_{\odot}$. Our current model can not exclude the 10 keV WDM model.


\end{itemize}

Here we briefly discuss the limitation and implication of our results. Our model has two important inputs to describe the evolution of the inner structure of satellite galaxy. One is that the inner density slope is modified from initial NFW ($\alpha=1$) to a flat one by the baryonic feedback in the galaxy. The other is the tidal effect on the redistribution of the inner mass of satellite. Both descriptions have their limitation and uncertainties. For the first one, while some state-of-the-art hydrodynamical simulations, such as MaGICC, NIHAO and FIRE, have found that the density slope is correlated with the star formation efficiency, other studies like APOSTLE have not found the creation of a flat profile in satellite galaxy by baryonic feedback. The difference is contributed by a few factors, such as the subgrid physics implemented in these simulations, the star formation history or even the nature of dark matter. For more discussion on this issue, readers are referred to \cite{Bose19}. 

Once baryonic feedback is inefficient to flatten the density profile in model satellites, one must consider how to lower their circular velocities, especially for the satellite Carina, CVnI and Sextans. One possibility is that these satellites are outliers and they could form in haloes with lower concentration. One immediate consequence is whether the low-concentration subhaloes are able to survive the strong tidal force of the host halo. Another possibility is related to the second important ingredient of our model that tidal process can modify their inner mass distribution. For this purpose, we use the simulation results of \cite{Penarrubia10} to describe the effect of tidal heating. In their model it is found that the effect of tides on subhalo mass distribution is solely controlled by the total stripped mass fraction. The mass stripping depends on the orbit of the satellite galaxy. In our model we do not have the phase-space information of each subhalo but using an orbit-average mass loss rate for all subhaloes. In reality the mass stripping in the satellites could be different, leading to uncertainty on the predicted circular velocity.



Nevertheless, one robust conclusion of our model is that in the CDM, one must need tidal heating to lower the circular velocities of some satellite galaxies, such as Carina, CVnI and Sextans, to match the data. One will then expect to see the tidal feature in these satellites. As mentioned before, tidal stirring feature is clealy seen in Carina \citep{Lokas12}, but is uncertain in Sextans \citep{Roderick16}. As mentioned in these studies, unlike the satellite in the process of distinct disruption, such as Sagittarius, tidal features in normal low-surface dwarfs are very weak and has to be distinguished with their intrinsic bar structures. Current constraints are mainly from the optical data which are not deep enough. More observational data, including both optical and radio, are required to identify the tidal feature in the satellite galaxies, so as to quantify the extent of tidal heating on the inner structure of the satellite galaxies.



\section{Acknowledgments}

We thank Shaun Cole for kind help on the Monte-Carlo merger tree code in the WDM case. We also thank the anonymous referee for constructive and thoughtful suggestions to improve the paper. The work is supported  by the  973 program  (No.  2015CB857003), the NSFC (No. 11825303, 11861131006, 11333008).

\bibliographystyle{mn2e}
\bibliography{biblio}





\bsp	
\label{lastpage}
\end{document}